\documentclass[11pt]{article}

\usepackage[margin=1in]{geometry}
\usepackage{graphicx} 
\usepackage{amsmath} 

\usepackage{amssymb} 
\usepackage{mathrsfs} 
\usepackage{bm} 

\usepackage[font=footnotesize,labelfont=bf,labelsep=space]{caption} 
\usepackage{authblk} 
\usepackage[squaren]{SIunits} 
\usepackage{hyperref} 
\usepackage{cite} 
\usepackage{tensor} 

\DeclareMathAlphabet{\mathpzc}{OT1}{pzc}{m}{it}

\usepackage{fixfoot}

\DeclareFixedFootnote{\rep}{Both authors contributed equally to this work.}

\begin{document}

\title{Linked and Knotted Gravitational Radiation}

\author[1,2]{Amy Thompson\rep{}\footnote{email: amy@physics.ucsb.edu}}
\author[2,3]{Joe Swearngin $^*$}
\author[1,2]{Dirk Bouwmeester}

\affil[1]{\small Dept. of Physics, University of California, Santa Barbara, CA 93106}
\affil[2]{\small Huygens Laboratory, Leiden University, PO Box 9504, 2300 RA Leiden, The Netherlands}
\affil[3]{\small Dept. of Physics, University of California, Los Angeles, CA 90095}

\date{}

\maketitle

\begin{abstract}
We show that the torus knot topology is inherent in electromagnetic and gravitational radiation by constructing spin-$N$ fields based on this topology from the elementary states of twistor theory. The twistor functions corresponding to the elementary states admit a parameterization in terms of the poloidal and toroidal winding numbers of the torus knots, allowing one to choose the degree of linking or knotting of the associated field configuration. Using the gravito-electromagnetic formalism, we show that the torus knot structure is exhibited in the tendex and vortex lines for the analogous linearized gravitational solutions. We describe the topology of the gravitational fields and its physical interpretation in terms of the tidal and frame drag forces of the gravitational field.   
\end{abstract}




\section{Introduction}

Knots and links are quite remarkable given that they are as old and ubiquitous as ropes and thread and yet have only relatively recently seen a rigorous formulation within mathematics. The study of knots and links has enjoyed a close relationship with physics since its inception by Gauss \cite{Przytycki2007history}. Today the application of these topological structures in theoretical physics is more widespread than it has ever been, from fault resistant quantum computing \cite{Nayak2008}, hadron models \cite{Skyrme1962,Faddeev1997}, topological MHD and fluid mechanics \cite{Kamchatnov1982,Thompson2014plasma}, classical field theories \cite{Ranada2002,Irvine2008,Swearngin2013}, quantum field theory \cite{Witten1989,Robertson1989}, DNA topology \cite{Arsuaga2005}, to nematic liquid crystals \cite{Machon2013} just to name a few. In this article we shall focus on the application of an important class of knots, torus knots, to classical electromagnetic and gravitational radiation.

A \emph{hopfion} is a field configuration based on a  topology derived from the Hopf fibration. The electromagnetic hopfion (EM hopfion) is a null solution to the source free Maxwell equations such that any two field lines associated to either the electric, magnetic, or Poynting vector fields (EBS fields) are closed and linked exactly once \cite{Ranada2002}. When an EM hopfion is decomposed onto hyperplanes of constant time there always exists a hyperplane wherein the EBS fields are tangent to the fibers of three orthogonal Hopf fibrations. If one extends the Poynting vector to be a future pointing light-like 4-vector its integral curves comprise a space filling shear-free null geodesic congruence dubbed the Robinson congruence by Roger Penrose. 

Ra{\~n}ada \cite{Ranada1989} rediscovered the EM hopfion solution and noted that its topology was invariant under time evolution. The search for generalizations of the hopfion solution led to the introduction of a set of non-null EM solutions based on torus knots \cite{Trueba2011}, but the topology was not preserved during time evolution. Around the same time, the Kerr-Robinson theorem was used to derive the hopfion from the Robinson congruence itself using 2-spinor methods \cite{Dalhuisen2012}. Inspired by the role of the Robinson congruence in the evolution of the hopfion, the conservation of field line topology was tied to the shear-free property of the Robinson congruence \cite{irvine2010linked}. 

\section{Complex Analytic and Twistor Methods}

The fundamental role of the Robinson congruence and its connection to the topology of physical systems leads one to consider its relationship to twistor theory. It has long been known that complex contour integral transforms can be used to find the solution to real PDEs. In 1903, Whittaker used this technique to construct the general solution to Laplace's equation \cite{Whittaker1904}. In 1915, Bateman extended this method to give solutions to the vacuum Maxwell equations \cite{Bateman1915}. Twistor theory was developed by Roger Penrose in the late 1960's as an extension of the $sl(2,\mathbb{C})$ spinor algebra. From this perspective, the complex analytic structure of Bateman can be related to the geometry of spinor fields on space-time \cite{Penrose1975}, which encode linear and angular momentum and are represented by two $SL(2,\mathbb{C})$ spinors $\pi_{A'}$ and $\omega ^{A}$. The linear momentum is the flagpole of $\pi_{A'}$ so that $p^a = \pi^{A'}\bar{\pi}^{A}$ and the angular momentum bivector is related to $\omega ^{A}$ by
\begin{equation*}
M^{ab} = i\omega ^{(A}\overline{\pi }^{B)} \epsilon^{A'B'} + c.c.
\end{equation*}
These spinors are combined into a single object $Z^{\alpha }=\left( \omega ^{A},\pi _{A'}\right)$ called a twistor. In this formalism, massless linear relativistic fields are expressed in the form of symmetric spinor fields. In 1969, the general solution to the massless spin-$N$ field equations was given as a complex contour integral transform now called the Penrose transform \cite{Penrose1969solutions}.

Within the twistor framework the solutions of the massless spin-$N$ equations are represented by the Penrose transform of homogeneous twistor functions (see Appendix). The elementary states are a canonical example of such functions whose singularities define Robinson congruences on Minkowski space $\mathbb{M}$. The space-time fields corresponding to the elementary states are finite-energy, and in the null case are everywhere non-singular \cite{Penrose1987origins}. For integer spin fields, the expansion of a solution over the elementary states in twistor space $\mathbb{T}$ is related to the expansion over spherical harmonics in $\mathbb{M}$ through the Penrose transform \cite{Grgin1966thesis}. These properties have made the elementary states the topic of many studies \cite{Penrose1972,Hughston1979advances,Hodges1982diagrams}, and for many problems it is assumed that considering the elementary states is sufficient to describe any solution \cite{Eastwood1991density}.

While investigating these twistor functions and their connection to field topology in $\mathbb{M}$, we have previously shown that the EM hopfion and the analogous gravitational hopfion are elementary states of twistor theory \cite{Swearngin2013}. Using the earlier construction for EM fields by Bateman \cite{Bateman1915}, Kedia, \emph{et al.} have shown that the EM hopfion is the simplest case in a set of null EM fields based on torus knots \cite{Irvine2013}. Here we show that field configurations based on all the torus knots are contained within the elementary states of twistor theory. The Hopf fibration appears as the degenerate case whereby the linked and knotted toroidal structure degenerates down to the linked hopfion configuration. This generalization leads to a construction for spin-$N$ fields based on torus knots. We will focus our analysis on the spin-1 and spin-2 fields, where the topology is physically manifest in the field lines.\footnote{For the Weyl fields, the linked and knotted topology appears in the current.}

The concept of tendex and vortex lines, gravitational lines of force for a particular observer, was developed by Nichols, \emph{et al.} \cite{Nichols2011}, who were motivated by the desire to understand the non-linear dynamics of curved space-time in a more intuitive, directly physical way than previous approaches. The physical understanding of the electromagnetic field is based upon the decomposition of the Faraday field strength tensor onto hyperplanes of constant time yielding two spacial vector fields interpreted as the electric and magnetic fields. Analogously, the Weyl curvature tensor admits a decomposition onto constant time hyperplanes yielding two spacial tensors called the gravito-electric (GE) and gravito-magnetic (GM) tensors.  The integral curves of the eigenvector fields of these tensors are called tendex and vortex lines respectively and represent the gravitational analog of electromagnetic field lines.  This method was elucidated through a series of papers where it was applied to I) weak field solutions \cite{Nichols2011}, II) stationary black holes \cite{Zhang2012}, and III) weak perturbations of stationary black holes \cite{Nichols2012}. This method of GEM decomposition is well-suited to studying linked and knotted fields because, as we have shown previously \cite{Swearngin2013}, the field topology is manifest in the lines of force for both the electromagnetic and analogous gravitational solutions.

\section{Parameterization of the Elementary States}

We will now relate the torus knot topology to the twistor elementary states to obtain solutions to the EM and gravitational spinor field equations. 

\begin{figure}[htb!]
\centering
 \includegraphics[width=1.0\textwidth]{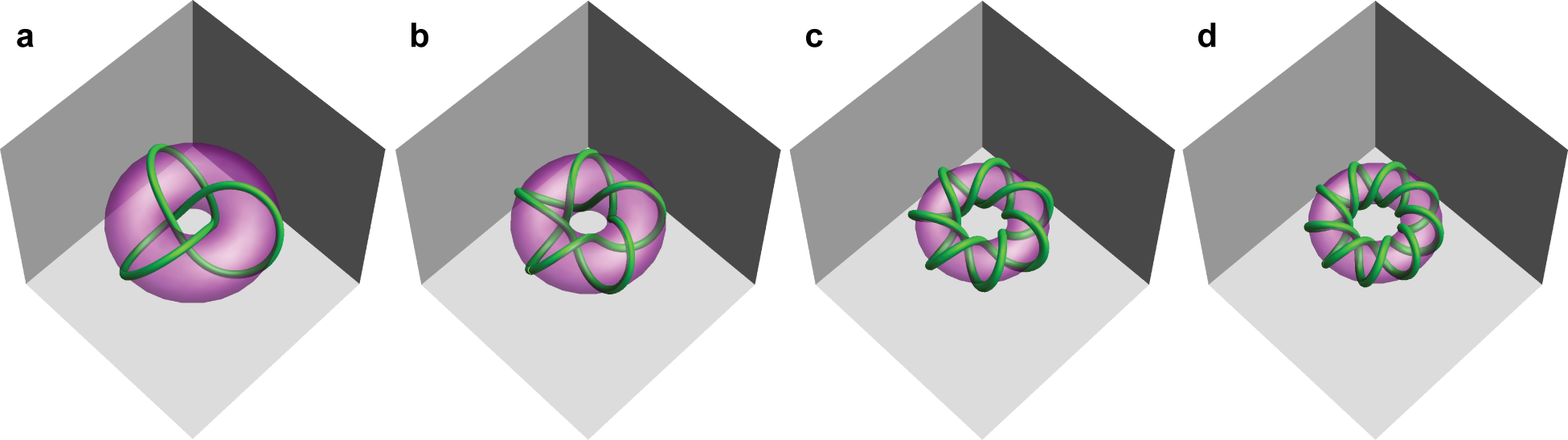}
 \caption{\label{fig:core_lines}Torus knots (green) wind $(n_t,n_p)$ times around a torus (purple) in the toroidal and poloidal directions, respectively. Shown here are the cases of a) trefoil (2,3) knot, b) cinquefoil (2,5) knot, c) septafoil (2,7) knot, and d) nonafoil (2,9) knot.}
\end{figure}

Torus knots are closed curves on the surface of a torus which wind an integer number of times about the toroidal direction $n_{t}$ and poloidal direction $n_{p}$ as in Fig. \ref{fig:core_lines}, where $n_t$ and $n_p$ are coprime and both greater than one. If $n_t$ and $n_p$ are not coprime, then there are $n_g=gcd(n_t,n_p)$ linked curves, each corresponding to a $(n_t,n_p)\bmod{n_g}$ torus knot. If either $n_t$ or $n_p$ is equal to one, then the knot is trivial with $n_g$ linked curves.

Following the twistor program, we represent solutions to the massless spin-$N$ equations in $\mathbb{M}$ as the Penrose transform of functions $f(Z)$ in twistor space \cite{PenroseSpinors2}. The details of the Penrose transform calculation for these fields are given in the Appendix.

Consider the twistor functions corresponding to the elementary states \cite{Penrose1972}
\begin{equation}
\label{eqn:twsitor_function}
f(Z)= \frac{(\bar C_{\gamma} Z^{\gamma})^{c}(\bar D_{\delta} Z^{\delta})^{d}}{(\bar A_{\alpha} Z^{\alpha})^a(\bar B_{\beta} Z^{\beta})^b}
\end{equation}
where $(\bar A_{\alpha} Z^{\alpha})$ is the $SU(2,2)$ twistor inner product. Choosing $a=1$ yields null/Type N solutions and we must have $b=2h+1+c+d$ to give the correct homogeneity $\mathpzc{h}=-2h-2$ for a solution with helicity $h$. We will show that the class of generating functions of the form\footnote{We use the conventions given in Eqn. \eqref{eqn:Pauli_matrices} of the Appendix.}
\begin{equation}
\label{eqn:generating_function}
f(Z)= \frac{(\bar C_{\gamma} Z^{\gamma})^{h(n_{p}-1)}(\bar D_{\delta} Z^{\delta})^{h(n_{t}-1)}}{(\bar A_{\alpha} Z^{\alpha})(\bar B_{\beta} Z^{\beta})^{h(n_{p}+n_{t})+1}},
\end{equation}
lead to field configurations with a torus knot topology where $n_{p}$ and $n_{t}$ correspond to the poloidal and toroidal winding numbers.

We choose the dual twistors
\begin{align}
\bar{A}_\alpha &= \imath (0,\sqrt{2},0,1)\notag\\
\bar{B}_\beta &= \imath (-\sqrt{2},0,-1,0)\notag\\
\bar{C}_\gamma &= (0,-\sqrt{2},0,1)\notag\\
\bar{D}_\delta &= \imath (-\sqrt{2},0,1,0). \label{eqn:dual_twistors}
\end{align}
$\bar{A}_\alpha$ and $\bar{C}_\gamma$ correspond to Robinson congruences with opposite twist, both with central axes aligned along the $+\hat z$-direction. $\bar{B}_\beta$ and $\bar{D}_\delta$ correspond to Robinson congruences with opposite twist, but in the $-\hat z$-direction. This choice leads to spin-$N$ fields which propagate $+\hat z$-direction with field line configurations that are based on a torus knot structure.

\section{Electromagnetic Torus Knots}

After applying the spin-1 Penrose transform to Eqn. (\ref{eqn:generating_function}), the resulting spinor field is
\begin{equation}
\label{spin1solution}
\phi_{A'B'}(x) = \frac{(\mathcal{A}_{C'}\mathcal{C}^{C'})^{n_{p}-1}(\mathcal{A}_{D'}\mathcal{D}^{D'})^{n_{t}-1}}{(\mathcal{A}_{E'}\mathcal{B}^{E'})^{n_{p}+n_{t}+1}}\mathcal{A}_{A'}\mathcal{A}_{B'}
\end{equation}
 Note that the Latin script spinor variables are the spinors associated to the Latin twistor variable. Ergo, $\mathcal{A}_{A'}$ is defined implicitly by $\bar A_{\alpha}Z^{\alpha} = \mathcal{A}_{A'} \pi^{A'}$ (see Appendix). The solution in Eqn. (\ref*{spin1solution}) satisfies the source-free spinor field equation by construction and yields the field strength spinor
\begin{align*}
\nabla^{AA'}\varphi_{A'B'} &= 0, \\
F_{A'B'AB} &= \varphi_{A'B'}\epsilon_{AB} + c.c. 
\end{align*}

\begin{figure}[htb!]
 \centering
 \includegraphics[width=1.0\textwidth]{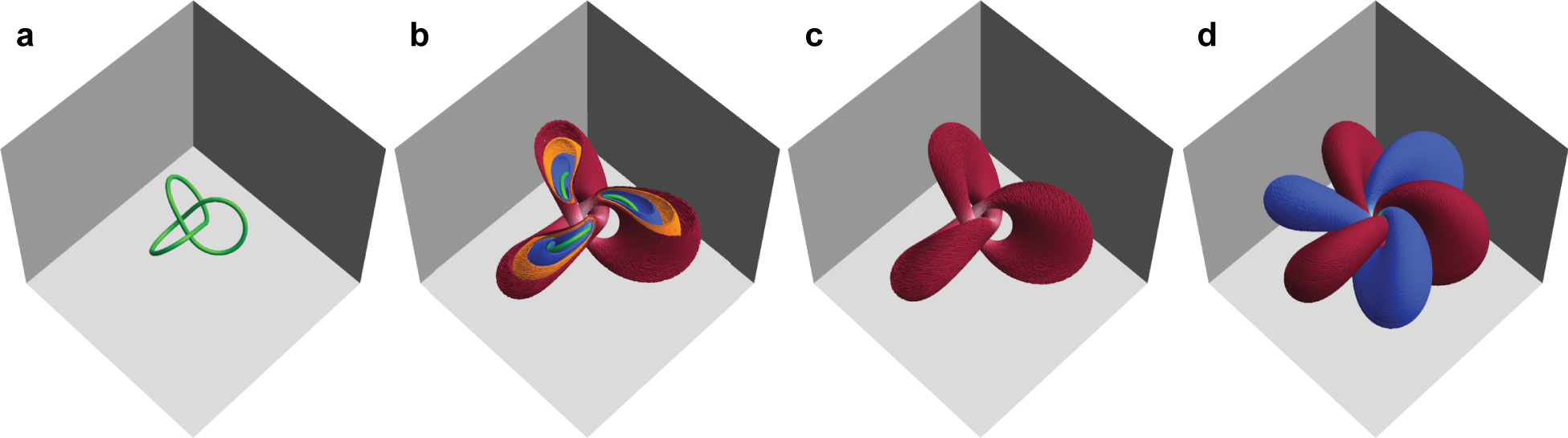}
 \caption{\label{fig:knot23surfaces}The field line structure based on (2,3) trefoil knot. a) The core field line is a torus knot (green). b) Each field line except the core lies on the surface of a nested, deformed torus. c) One field line fills a complete surface (red). d) Another field line fills a second surface (blue) linked with the first. The two linked core field lines and the nested surfaces around them fill all of space.}
\end{figure}

The spin-1 fields are null torus knots with a Poynting vector that is everywhere tangent to a Hopf fibration and propagates in the $\hat z$-direction without deformation. The solutions have the same topology\footnote{There is an overall constant factor of $4 n_t n_p$ in Ref. \cite{Irvine2013} that does not appear in our construction, but it does not affect the topology.} as the electromagnetic fields in Ref. \cite{Irvine2013}. The electric and magnetic vector fields each have the following topological structure as shown in Fig. \ref{fig:knot23surfaces}. There are $2n_g$ core field lines, where $n_g=gcd(n_t,n_p)$, which are linked (and knotted if $n_t,n_p>1$). Each core line has the same configuration as the corresponding torus knot with $(n_t,n_p)$ shown in Fig. \ref{fig:core_lines}. With the choice for $C$ and $D$ given in Eqn. (\ref{eqn:dual_twistors}), the poloidal and toroidal winding numbers for the EM case are related to the exponents in Eqn. (\ref{eqn:twsitor_function}) by $c=n_p-1$ and $d=n_t-1$. A single core field line is surrounded by nested, toroidal surfaces, each filled by one field line. A second core field line, also surrounded by nested surfaces, is linked with the first so that there are $2n_g$ sets of linked nested surfaces which fill all of space. A complete solution to Maxwell's equations consists of an electric and a magnetic field orthogonal to each other, both with this field line structure. The (1,1) case corresponds to the electromagnetic hopfion.

\begin{figure}[htb!]
 \centering
 \includegraphics[width=0.8\textwidth]{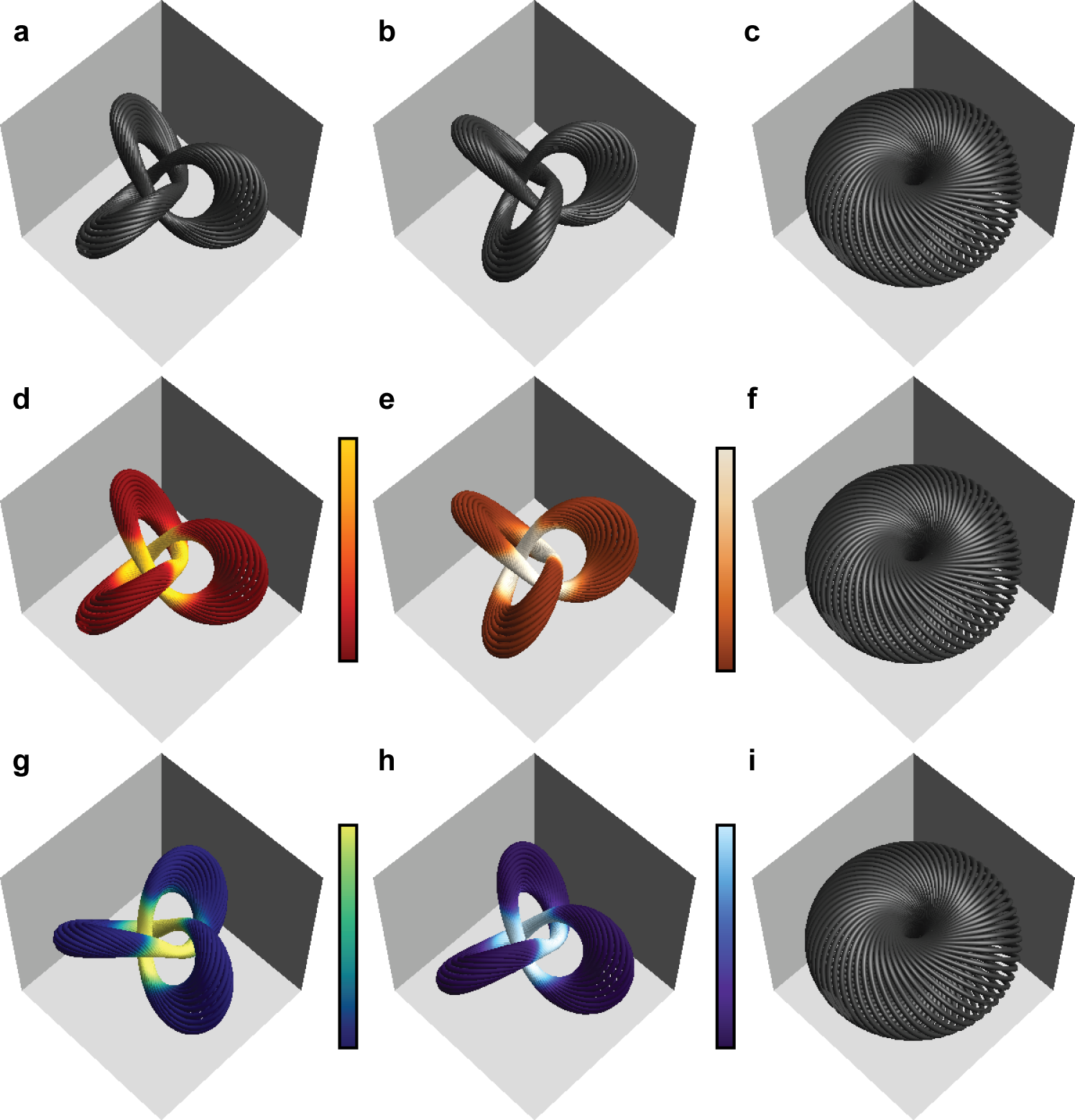}
 \caption{A comparison of the spin-1 (EM) and spin-2 (gravity) trefoil knots at $t=0$. The first row is the EM trefoil knot: \textbf{a} the electric field, \textbf{b} the magnetic field, and \textbf{c} the Poynting vector field. The second row is the gravito-electric trefoil knot: \textbf{d} the negative eigenvalue field $\vec e_-$, \textbf{e} the positive eigenvalue field $\vec e_+$, and \textbf{f} the zero eigenvalue field $\vec e_0$.The third row is the gravito-magnetic trefoil knot: \textbf{g} the negative eigenvalue field $\vec b_-$, \textbf{h} the positive eigenvalue field $\vec b_+$, and \textbf{i} the zero eigenvalue field $\vec b_0$. The color scale indicates magnitude of the eigenvalue, with lighter colors indicating a higher magnitude.}
\label{fig:knot23}
\end{figure}

At $t=0$ the electric and magnetic fields are tangent to orthogonal torus knots, as shown in Fig. \ref{fig:knot23} (first row). The fields will deform under time evolution, but the topology will be conserved since $\vec E \cdot \vec B=0$ \cite{irvine2010linked,Arrayas2011exchange}. 


\section{Gravito-electromagnetic Torus Knots}
The spin-2 solutions will be analyzed in terms of the gravito-electromagnetic tidal tensors. The Weyl tensor $C_{abcd}$ can be decomposed into an even-parity ``electric'' part $E_{ij}$ corresponding to the tidal field and an odd-parity ``magnetic'' part $B_{ij}$ for the frame-drag field, in direct analogy with the decomposition of the electromagnetic field strength tensor into an electric field and a magnetic field. For an observer at rest, this gives
\begin{align}
E_{ij} &= C_{i0j0} \\
B_{ij} &= - \ast C_{i0j0}.
\end{align}
These tensors are symmetric and traceless, and are thus characterized entirely by their eigenvalues and eigenvectors. One may then study the eigensystem associated with these matrix-valued fields instead of the Weyl tensor itself, allowing for a more intuitive approach to understanding the gravitational field. The integral curves of the eigenvectors of the tidal tensor are called tendex lines and their eigenvalues define the tidal acceleration along these lines. The integral curves of the eigenvectors of the frame-drag tensor are called vortex lines and their eigenvalues define the gyroscope precession about the vortex lines. Together, the tendex and vortex lines are the analog of electromagnetic field lines.\cite{Nichols2011}

After applying the spin-2 Penrose transform to Eqn. (\ref{eqn:generating_function}), the resulting spinor field is
\begin{equation}\phi_{A'B'C'D'}(x) = \frac{(\mathcal{A}_{F'}\mathcal{C}^{F'})^{2(n_{p}-1)}(\mathcal{A}_{G'}\mathcal{D}^{G'})^{2(n_{t}-1)}}{(\mathcal{A}_{E'}\mathcal{B}^{E'})^{2(n_{p}+n_{t})+1}}\mathcal{A}_{A'}\mathcal{A}_{B'}\mathcal{A}_{C'}\mathcal{A}_{D'}.
\end{equation}
The source-free field equation and Weyl field strength spinor are 
\begin{align*}
\nabla^{AA'}\varphi_{A'B'C'D'} &= 0, \\ 
C_{A'B'C'D'ABCD} &= \varphi_{A'B'C'D'}\epsilon_{AB}\epsilon_{CD} + c.c. 
\end{align*}

The Weyl tensor can then be decomposed into the GEM components. For Type N, the eigenvalues for both the GE and GM tensors take the form $\{ -\Lambda, 0, +\Lambda \}$, with $-\Lambda(x) \leq 0 \leq +\Lambda(x)$ for all points $x$ in space-time The magnitude of the eigenvalues is
\begin{equation}
\label{eqn:eigenvalue_Lambda}
\Lambda=\frac{2^{2n_p-3}(1 + r^2 + t^2 -2tz)^2(r^2 - z^2)^{n_p - 1}(r^4 - 2 r^2 (1 + t^2) + (1 + t^2)^2 + 4 z^2)^{n_t - 1}}{(r^4 - 2 r^2 (-1 + t^2) + (1 + t^2)^2)^{\frac{5}{2} + n_t + n_p}}
\end{equation}

\begin{figure}[htb!]
 \centering
 \includegraphics[width=1.0\textwidth]{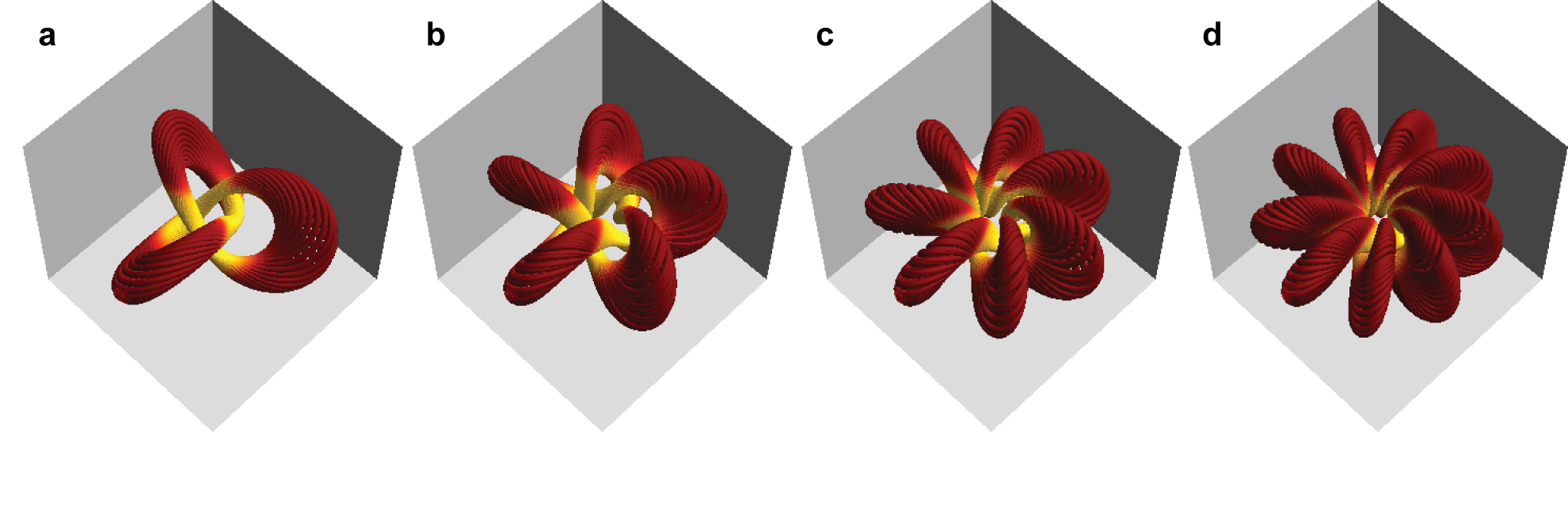}
 \caption{The eigenvector field $\vec e_-$ for the gravitational field based on the a) trefoil (2,3) knot, b) cinquefoil (2,5) knot, c) septafoil (2,7) knot, and d) nonafoil (2,9) knot. The color scaling is the same as in Fig. \ref{fig:knot23}.}
 \label{fig:Eminus_t0}
\end{figure}

We label the eigenvectors $\{\vec{e}_-, \vec{e}_0, \vec{e}_+\}$ and $\{\vec{b}_-, \vec{b}_0, \vec{b}_+\}$ corresponding to the eigenvalues for the tidal and frame-drag fields respectively. For the zero eigenvalue, the eigenvectors $\vec{e}_0$ and $\vec{b}_0$ are both aligned with the Poynting vector of the null EM torus knots. For the remaining eigenvectors, we can construct Riemann-Silberstein (RS) vectors $\vec{f}_e = \vec{e}_- + i\vec{e}_+$ and $\vec{f}_b = \vec{b}_- + i\vec{b}_+$ which are related to each other by
\begin{equation}
\label{eqn:RS_relation}
\vec{f}_e = e^{i\pi/4} \vec{f}_b.
\end{equation}
At $t=0$, the eigenvectors of the GE fields have precisely the same structure as the EM fields, and the GM eigenvector fields have the same structure but rotated by $45^\circ$. For the spin-$2$ case, the poloidal and toroidal winding numbers are related to the exponents in Eqn. (\ref{eqn:twsitor_function}) by $c=2(n_p-1)$ and $d=2(n_t-1)$. The surfaces of the $\vec{e}_-$ eigenvector, color-scaled according to the magnitude of the eigenvalue, for different values of $(n_t,n_p)$ are shown in Fig. \ref{fig:Eminus_t0}. The other GEM fields can be constructed by rotating $\vec{e}_-$ according to Eqn. (\ref{eqn:RS_relation}): $\vec{e}_+$ is found by rotating $\vec{e}_-$ by $90^\circ$ about the Poynting vector. $\vec{b}_-$ and $\vec{b}_+$ are found by rotating $\vec{e}_-$ and $\vec{e}_+$ by $45^\circ$, respectively. The eigenvalues of the GEM fields for a given $(n_t,n_p)$ have the same magnitude (color-scaling) given by $|\Lambda(x)|$ in Eqn. (\ref{eqn:eigenvalue_Lambda}).

\section{Conclusion}

Here we have shown that the null EM torus knot solutions correspond to a class of elementary states characterized the poloidal and toroidal winding numbers of the associated field configuration. Using the relationship between fields of different spin in the twistor formalism, we constructed the analogous gravitational radiation configuration that possesses tendex and vortex lines based on a torus knot structure. Since the topology is manifest in the tendex and vortex lines, the gravito-electromagnetic tidal tensor decomposition is a straightforward method for characterizing these field configurations.

The elementary states were known as early as the 1970's \cite{Penrose1972}, however the explicit forms of their associated spinor and tensor representations on $\mathbb{M}$ were never published.\footnote{from private discussions with Roger Penrose} The modern rediscovery of these solutions has raised interest in obtaining a more complete physical understanding of the topological properties of these fields. The parameterization of the twistor functions corresponding to the elementary states in terms of the poloidal and toroidal winding indicates that the torus knot structure is indeed inherent in the elementary states. For both electromagnetism and gravity, the topology appears in the configuration of the lines of force.

\section*{Acknowledgments}

The authors would like to thank J.W. Dalhuisen for discussions and A. Wickes for help with the figures. This work is supported by NWO VICI 680-47-604 and NSF Award PHY-1206118.

\section*{Appendix: The Penrose Transform for Spin-$N$ Torus Knots}
\label{sec:transform}

To obtain a spinor field $\varphi _{A_{1}'\cdots A_{2h}'}\left( x\right)$ with helicity $h$ which satisfies the spin-$N$ massless field equation 
\begin{equation*}
\nabla ^{AA_{1}'}\varphi _{A_{1}'\cdots A_{2h}'}\left( x\right) = 0
\end{equation*}
we will calculate the Penrose transform 
\begin{equation}
\label{penroseTransform}
\varphi_{A'_1\cdots A'_{2h}}(x) = \frac{1}{2\pi i} \oint_\Gamma\pi_{A'_1}\cdots\pi_{A'_{2h}}f(Z)\pi_{B'}d\pi^{B'}
\end{equation}
where $\Gamma$ is a contour on the Celestial sphere of $x$ that separates the poles of $f(Z)$. Consider the twistor function given by Eqn. (\ref{eqn:generating_function})
\begin{equation}
f(Z)= \frac{(\bar C_{\gamma} Z^{\gamma})^{h(n_{p}-1)}(\bar D_{\delta} Z^{\delta})^{h(n_{t}-1)}}{(\bar A_{\alpha} Z^{\alpha})(\bar B_{\beta} Z^{\beta})^{h(n_{p}+n_{t})+1}}. \notag
\end{equation}
(For further review of the background material on twistors and the Penrose transform see Ref. \cite{Swearngin2013}). Let $\overline{A}_\alpha = (\mu_A,\lambda^{A'})$ be a dual twistor such that
\begin{align}
\overline{A}_\alpha Z^\alpha &= i\mu_A x^{AA'} \pi_{A'} + \lambda^{A'} \pi_{A'}  \notag \\ 
	&\equiv \mathcal{A}^{A'} \pi_{A'}
\end{align}
where $Z^\alpha = (ix^{AA'} \pi_{A'}, \pi_{A'})$. Similar relations hold for the other dual twistors $\overline{B}_\beta Z^\beta \equiv \mathcal{B}^{B'} \pi_{B'}$, $\overline{C}_\gamma Z^\gamma \equiv \mathcal{C}^{C'} \pi_{C'}$, and $\overline{D}_\delta Z^\delta \equiv \mathcal{D}^{D'} \pi_{D'}$.
We want to write the Penrose transform as an integral over the $\mathbb{CP}^1$ coordinate $\zeta = \pi_{1'} / \pi_{0'}$, so we have 
\begin{align}
\pi_{C'} d\pi^{C'} &= \pi_{C'} d\pi_{D'} \epsilon^{D'C'}  \notag \\
	&= \pi_{0'} d\pi_{1'} - \pi_{1'} d\pi_{0'}  \notag \\
	&= (\pi_{0'})^2 d(\frac{\pi_{1'}}{\pi_{0'}}).
\end{align}
Adopting the canonical spin bases $\{o_{A'}, \iota_{A'}\}$ we have that
\begin{align}
\pi_{A'} &= \pi_{0'} o_{A'} + \pi_{1'} \iota_{A'}  \notag \\
	&= \pi_{0'} (o_{A'} + (\frac{\pi_{1'}}{\pi_{0'}}) \iota_{A'}).
\end{align}
Observing that
\begin{align}
\frac{1}{\pi_{0'}} \mathcal{A}^{A'} \pi_{A'} &= \mathcal{A}^{0'} + \mathcal{A}^{1'} (\frac{\pi_{1'}}{\pi_{0'}})
\end{align}
and similarly for $\mathcal{B}$, $\mathcal{C}$, and $\mathcal{D}$, we see that the Penrose transform becomes an integral manifestly over $\mathbb{CP}^1$. Thus 
\begin{align}
\label{eqn:penroseTransformResult}
\varphi_{A'_1\cdots A'_{2h}}(x) &= \frac{1}{2\pi i} \oint_\Gamma f(Z)\pi_{A'_1} \cdots \pi_{A'_{2h}} \pi_{B'}d\pi^{B'}  \\
 &= \frac{1}{2\pi i} \oint_\Gamma \frac {(\mathcal{C}^{0'}+\mathcal{C}^{1'}(\frac{\pi_{1'}}{\pi_{0'}}))^{h(n_p-1)}(\mathcal{D}^{0'}+\mathcal{D}^{1'}(\frac{\pi_{1'}}{\pi_{0'}}))^{h(n_t-1)}  } {(\mathcal{A}^{0'}+\mathcal{A}^{1'}(\frac{\pi_{1'}}{\pi_{0'}}))(\mathcal{B}^{0'}+\mathcal{B}^{1'}(\frac{\pi_{1'}}{\pi_{0'}}))^{h(n_p+n_t)+1} } (o_{A_1}+(\frac{\pi_{1'}}{\pi_{0'}}) \iota_{A'_1}) \cdots (o_{A_{2h}}+(\frac{\pi_{1'}}{\pi_{0'}}) \iota_{A'_{2h}}) d(\frac{\pi_{1'}}{\pi_{0'}}) \notag \\
	&= \frac{(\mathcal{C}^{1'})^{h(n_p-1)}(\mathcal{D}^{1'})^{h(n_t-1)}}{2\pi i \mathcal{A}^{1'} (\mathcal{B}^{1'})^{h(n_p+n_t)+1}} \oint_\Gamma \frac{(\rho+\zeta)^{h(n_p-1)}(\tau+\zeta)^{h(n_t-1)}} {(\mu+\zeta)(\nu+\zeta)^{h(n_p+n_t)+1}} (o_{A_1}+\zeta\iota_{A'_1})\cdots(o_{A_{2h}}+\zeta\iota_{A'_{2h}}) d\zeta  \notag
\end{align}
where $\mu = \mathcal{A}^{0'} / \mathcal{A}^{1'}$, $\nu = \mathcal{B}^{0'} / \mathcal{B}^{1'}$, $\rho = \mathcal{C}^{0'} / \mathcal{C}^{1'}$, and $\tau = \mathcal{D}^{0'} / \mathcal{D}^{1'}$.

The above change of variables leaves the contour integral in a form that is straightforward to calculate. The contour $\Gamma$ is taken to enclose the pole $-\mu$ giving the result
\begin{align}
\varphi_{A'_1 \cdots A'_{2h}}(x) &= \frac{(\mathcal{C}^{1'})^{h(n_p-1)}(\mathcal{D}^{1'})^{h(n_t-1)}}{\mathcal{A}^{1'} (\mathcal{B}^{1'})^{h(n_p+n_t)+1}} \underset{\zeta=-\mu}{\text{Res}} \frac{(\rho+\zeta)^{h(n_p-1)}(\tau+\zeta)^{h(n_t-1)} } {(\mu+\zeta)(\nu+\zeta)^{h(n_p+n_t)+1}} (o_{A_1}+\zeta\iota_{A'_1})\cdots(o_{A_{2h}}+\zeta\iota_{A'_{2h}}) \notag \\
&= \frac{(\mathcal{C}^{1'})^{h(n_p-1)}(\mathcal{D}^{1'})^{h(n_t-1)}}{\mathcal{A}^{1'} (\mathcal{B}^{1'})^{h(n_p+n_t)+1}} \frac{(\rho -\mu)^{h(n_p-1)}(\tau-\mu)^{h(n_t-1)}} {(\nu-\mu)^{h(n_p+n_t)+1}} (o_{A'_1}-\mu\iota_{A'_1})\cdots(o_{A'_{2h}}-\mu\iota_{A'_{2h}}) \notag \\
&= \frac{(\mathcal{A}^{1'}\mathcal{C}^{0'} - \mathcal{A}^{0'}\mathcal{C}^{1'})^{h(n_p-1)}(\mathcal{A}^{1'}\mathcal{D}^{0'} - \mathcal{A}^{0'}\mathcal{D}^{1'})^{h(n_t-1)}} {(\mathcal{A}^{1'}\mathcal{B}^{0'} - \mathcal{A}^{0'}\mathcal{B}^{1'})^{h(n_p+n_t)+1}} (\mathcal{A}^{1'} o_{A'_1} - \mathcal{A}^{0'} \iota_{A'_1}) \cdots (\mathcal{A}^{1'} o_{A'_{2h}} - \mathcal{A}^{0'} \iota_{A'_{2h}}) \notag \\
&= \frac{(\epsilon_{C'D'}\mathcal{A}^{C'}\mathcal{C}^{D'})^{h(n_p-1)}(\epsilon_{E'F'}\mathcal{A}^{E'}\mathcal{D}^{F'})^{h(n_t-1)}}{(\epsilon_{A'B'}\mathcal{A}^{A'}\mathcal{B}^{B'})^{h(n_p+n_t)+1}}\mathcal{A}_{A'_1} \cdots \mathcal{A}_{A'_{2h}} \notag \\
&=
\frac{(\mathcal{A}_{C'}\mathcal{C}^{C'})^{h(n_p-1)}(\mathcal{A}_{D'}\mathcal{D}^{D'})^{h(n_t-1)}}{(\mathcal{A}_{B'}\mathcal{B}^{B'})^{h(n_p+n_t)+1}}\mathcal{A}_{A'_1} \cdots \mathcal{A}_{A'_{2h}}
\label{eqn:knotField}.
\end{align}
In the case of spin-1 and spin-2, the classical field strength spinors are given by
\begin{eqnarray}
F_{A_{1}'A_{2}'\ A_{1}A_{2}} &=&\varphi _{A_{1}'A_{2}'}\epsilon _{A_{1}A_{2}} + \overline{\varphi} _{A_{1}A_{2}}\epsilon _{A_{1}'A_{2}'} \notag \\
C_{A_{1}'\cdots A_{4}'\ A_{1}\cdots A_{4}} &=& \varphi
_{A_{1}'\cdots A_{4}'}\ \epsilon _{A_{1}A_{2}}\ \epsilon
_{A_{3}A_{4}}+\overline{\varphi}_{A_{1}\cdots A_{4}}\ \epsilon _{A_{1}'A_{2}'}\ \epsilon_{A_{3}'A_{4}'}.
\label{classicalFieldStrength}
\end{eqnarray}
Choosing the standard basis to be the extended Pauli matrices we define the following symbols, referred to as the \textit{Infeld-van der Waerden symbols}
\begin{center}
\begin{equation}
\begin{array}{cccc}
\sigma _{0}^{AA'}\equiv \frac{1}{\sqrt{2}}
\begin{pmatrix}
1 & 0 \\ 
0 & 1
\end{pmatrix}
& \sigma _{1}^{AA'}\equiv \frac{1}{\sqrt{2}}
\begin{pmatrix}
0 & 1 \\ 
1 & 0
\end{pmatrix}
& \sigma _{2}^{AA'}\equiv \frac{1}{\sqrt{2}}
\begin{pmatrix}
0 & i \\ 
-i & 0
\end{pmatrix}
& \sigma _{3}^{AA'}\equiv \frac{1}{\sqrt{2}}
\begin{pmatrix}
1 & 0 \\ 
0 & -1
\end{pmatrix}
.
\end{array}
\label{eqn:Pauli_matrices}
\end{equation}
\end{center}
The spinor fields are then related to the world tensor description by
\begin{eqnarray}
F_{ab} &=& F_{A_{1}'A_{2}'\ A_{1}A_{2}}\sigma _{a}^{A_1A_1'}\sigma _{b}^{A_2A_2'} \notag \\
C_{abcd} &=& C_{A_{1}'\cdots A_{4}'\ A_{1}\cdots A_{4}}\sigma _{a}^{A_1A_1'} \cdots \sigma _{d}^{A_4A_4'}.
\end{eqnarray}

\end{document}